\documentclass[twocolumn,showpacs,preprintnumbers,amsmath,amssymb,showkeys]{revtex4}
 
\usepackage{graphicx}

\newcommand{\be}{\begin{equation}}
  \newcommand{\ee}{\end{equation}}
\newcommand{\bea}{\begin{eqnarray}}
  \newcommand{\eea}{\end{eqnarray}}

\newcommand{\bml}{\begin{mathletters}}
  \newcommand{\eml}{\end{mathletters}}

\newcommand{\vecx}{{\bf x}}
\newcommand{\vecy}{{\bf y}}
\newcommand{\vecz}{{\bf z}}

\newcommand{\veceta}{{\bf \eta}}
\newcommand{\vecxsi}{{\bf \xi}}
\newcommand{\vecr}{{\bf r}}
\newcommand{\vecu}{{\bf u}}

\begin{document}

\title{Robustly estimating the flow direction of information in
  complex physical systems}

\author{Guido Nolte}\email{nolte@first.fraunhofer.de}
\affiliation{Fraunhofer FIRST.IDA, Kekul\'estrasse 7, D-12489  Berlin, Germany} 
\author{Andreas Ziehe}\email{ziehe@first.fraunhofer.de}
\affiliation{Technical University of Berlin, Computer Science, Machine Learning Laboratory, Franklinstr. 28/29, 10587 Berlin, Germany}

\author{Vadim V. Nikulin}\email{vadim.nikulin@charite.de}
\affiliation{Dept of Neurology, Campus Benjamin Franklin, Charite University Medicine Berlin, D-12203, Germany} 
\author{Alois Schl\"ogl}\email{alois.schloegl@first.fraunhofer.de}
\affiliation{Fraunhofer FIRST.IDA, Kekul\'estrasse 7, D-12489  Berlin, Germany \mbox{} } 
\author{Nicole Kr\"amer}\email{nkraemer@cs.tu-berlin.de}
\affiliation{Technical University of Berlin, Computer Science, Machine Learning Laboratory, Franklinstr. 28/29, 10587 Berlin, 
Germany \mbox{ }}
\author{Tom Brismar}
\email{tom.brismar@ki.se}
\affiliation{Karolinska Institutet, Clinical Neurophysiology, Karolinska Hospital, S-17176 Stockholm, Sweden}
\author{Klaus-Robert M\"uller}\email{krm@cs.tu-berlin.de}
\affiliation{Technical University of Berlin, Computer Science, Machine Learning Laboratory, Franklinstr. 28/29, 10587 Berlin, Germany and Fraunhofer FIRST.IDA, Germany}

\date{\today}

\begin{abstract}

  We propose a new measure to estimate the direction of information
  flux in multivariate time series from complex systems.  This
  measure, based on the slope of the phase spectrum (Phase Slope
  Index) has invariance properties that are important for applications
  in real physical or biological systems: (a) it is strictly
  insensitive to mixtures of arbitrary independent sources, (b) it
  gives meaningful results even if the phase spectrum is not linear,
  and (c) it properly weights contributions from different
  frequencies. Simulations of a class of coupled multivariate random
  data show that for truly unidirectional information flow without
  additional noise contamination our measure detects the correct
  direction as good as the standard Granger causality. For random {\em
    mixtures} of independent sources Granger Causality {\em
    erroneously} yields highly significant results whereas our measure
  {\em correctly} becomes non-significant.
  An application of our novel method to EEG data (88 subjects in
  eyes-closed condition) reveals a strikingly clear front-to-back
  information flow in the vast majority of subjects and thus
  contributes to a better understanding of information processing in
  the brain.
\end{abstract}


\maketitle

To understand interacting systems it is of fundamental importance to
distinguish the driver from the recipient, and hence to be able to
estimate the direction of information flow.  If one cannot interfere
with the system, the direction can be estimated with a temporal
argument: the driver is earlier than the recipient from which it
follows that the driver contains information about the future of the
recipient not contained in the past of the recipient while the reverse
is not the case.  This argument is the conceptual basis of Granger
Causality \cite{Gra69,Gra80} which is probably the most prominent
method to estimate the direction of causal influence in time series
analysis.  Granger Causality was originally developed in econometry,
but is applied to many different problems in physics, geosciences
(cause of climate change), social sciences, and biology with special
emphasis on neural system \cite{Kau97,Nar06,Mar06,Bro04,Sat06}.

The difficulty in realistic measurements in complex systems is that
asymmetries in detection power may as well arise due to other factors,
specifically independent background activity having nontrivial
spectral properties and eventually being measured in unknown
superposition in the channels. In this case the interpretation of the
asymmetry as a direction of information flow can lead to significant
albeit false results \cite{Alb04}.  The purpose of this paper is to
propose a novel estimate of flux direction which is highly robust
against false estimates caused by confounding factors of very general
nature.

More formally, we are interested in statistical dependencies in
complex physical systems and especially in causal relations between a
signal of interest consisting of two sources with time series $x_i(t)$
for $i=1,2$.  The measured data $\vecy(t)$ are assumed to be a
superposition of these sources of interest and additive noise
$\veceta(t)$ in the form \be \vecy(t)=\vecx(t)+B\veceta(t)
\label{data} \ee where $\veceta(t)$ is a set of $M$ independent noise
sources which are mixed into the measurement channels by an unknown
$2\times M$ mixing matrix $B$.

The proposed method is based on the slope of the phase of
cross-spectra between two time series.  A fixed time delay for an
interaction between two systems will affect different frequency
components in different ways. This is most easily seen if we assume
that the interaction is merely a delay by a time $\tau$, i.e.
$y_2(t)=ay_1(t-\tau)$ with $a$ being some constant.  In the
Fourier-domain this relation reads $\hat{y}_2(f)=a\exp(-i2\pi
f\tau)\hat{y}_1(f)$. For the cross-spectrum $S_{ij}(f)$ between the
two channels $i$ and $j$ one has \be S_{12}(f)=\langle
\hat{y}_1(f)\hat{y}_2^*(f)\rangle \sim \exp(i2\pi
f\tau)\equiv\exp(i\Phi(f))
 \label{S}
 \ee
 where $\langle \cdot \rangle$ denotes expectation value. The phase-spectrum $\Phi(f)=2\pi f\tau$ 
 is linear and proportional
 to the time delay $\tau$. The slope of $\Phi(f)$ can be estimated, 
 and the causal direction is estimated to go from $y_1$ to $y_2$ ($y_2$ to $y_1$) if it is positive
(negative).

The idea here is now to define an average phase slope in such a way 
that a) this quantity properly represents relative time delays of 
different signals and especially coincides with the classical definition 
for linear phase spectra , b) it is insensitive to signals which do 
not interact regardless of spectral content and superpositions of these 
signals, and c) it properly weights different frequency regions according 
to the statistical relevance. This quantity is termed  'Phase Slope Index' (PSI)
and is defined as
\be \tilde{\Psi}_{ij}=\Im\left(
\sum_{f \in F} C_{ij}^*(f)C_{ij}(f+\delta f)\right) \ee 
where
\be
 C_{ij}(f)=\frac{S_{ij}(f)}{\sqrt{S_{ii}(f)S_{jj}(f)}}
 \ee
is the complex coherency, $S$ is the cross-spectral matrix, $\delta f$ is the frequency 
 resolution, and $\Im(\cdot)$ denotes taking the imaginary part.
  $F$ is the set of frequencies
 over which the slope is summed. 
   
 To see that the definition of $\tilde{\Psi}_{ij}$ corresponds to a
 meaningful estimate of the average slope it is convenient to rewrite
 it as \be \tilde{\Psi}_{ij}=\sum_{f\in F}
 \alpha_{ij}(f)\alpha_{ij}(f+\delta f) \sin(\Phi(f+\delta f)-\Phi(f))
 \ee with $\alpha_{ij}(f)=|C_{ij}(f)|$ being frequency dependent
 weights.  For smooth phase spectra, $\sin(\Phi(f+\delta
 f)-\Phi(f))\approx \Phi(f+\delta f)-\Phi(f)$ and hence $\Psi$
 corresponds to a weighted average of the slope.  We emphasize that
 since $\Psi$ vanishes if the imaginary part of coherency vanishes it
 will be {\em insensitive} to mixtures of non-interacting sources
 \cite{NolBai04,nolte06}.

 Finally, it is convenient to normalize $\tilde{\Psi}$ by an estimate
 of its standard deviation \be
 \Psi=\frac{\tilde{\Psi}}{std(\tilde{\Psi})} \ee with
 $std(\tilde{\Psi})$ being estimated by the Jackknife method.  In the
 examples below we always show normalized measures of directionality,
 and we consider absolute values larger than 2 as significant.

 Estimations of cross-spectra is standard \cite{nunez97,NolBai04} but
 technical details may differ. Here, we first divide the whole data into
 epochs containing continuous data (4 seconds duration), then we
 divide each epoch further into segments of time $T$, here of 2
 seconds duration corresponding to a frequency resolution of $\delta
 f=0.5$ Hz, multiply the data for each segment with a Hanning window,
 Fourier-transform the data, and estimate the cross-spectra according
 to Eq.\ref{S} as an average over all segments.  The segments have
 50\% overlap within each epoch but not across epochs.  To apply the
 Jackknife method, for each pair of channels we calculate
 $\tilde{\Psi}_k$ from data with the $k.th$ epoch removed for all $k$.
 The standard deviation of $\tilde{\Psi}$ is finally estimated for $K$
 epochs as $\sqrt{K}\sigma$ where $\sigma$ is the standard deviation
 of the set of $\tilde{\Psi}_k$.
 
 Our new method is compared to Granger causality using Autoregressive
 (AR) models both for wide band and narrow band analysis \cite{Din06}
 with analogous normalization by the estimated standard deviation.  To
 estimate the parameters of the model we here use the
 Levinson-Wiggens-Robinson \cite{marple87} algorithm available in the
 open Biosig toolbox \cite{biosig}.  Granger Causality is defined as
 the difference between the flux from channel $1$ to $2$ and the flux
 from channel $2$ to $1$ normalized to unit estimated standard
 deviation.

\begin{figure}[ht]
\centering\resizebox*{8cm}{7cm}{{\includegraphics{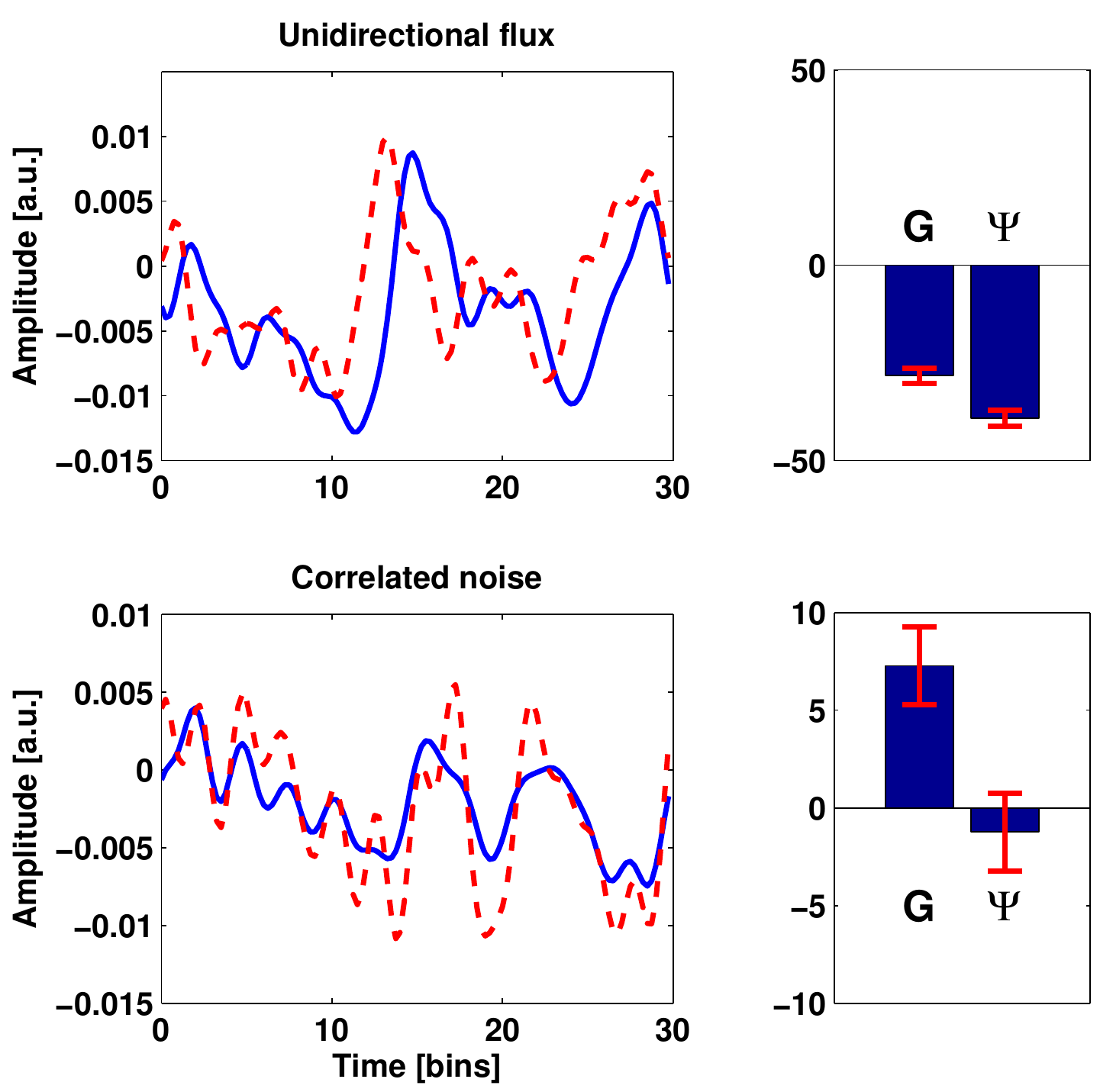}}}
\caption{Upper panels: strong interaction from second (red in left panels) to first (blue in left panels) 
signal. Lower 
panels: mixture of brown and white noise. The error bars in the right panels 
indicate estimated 95\% error margins corresponding to 2 standard deviations. 
Time series in the left panels were upsampled to 400 Hz. 
\label{fig_simexample}}
\end{figure}

 We first illustrate typical results for two simple cases in
 Fig.\ref{fig_simexample}. The upper panels show a simulation of a
 strong interaction from the second (dashed) to the first channel
 (solid) generated with a simple AR model of order one. The second
 signal is clearly earlier than the first signal. Both methods detect
 this direction correctly from only 2000 data points. In the lower
 panels we show a mixture of pink and white noise. In contrast to
 PSI, Granger causality erroneously still detects a significant direction.

To study a more general class of signals we simulated data with
structure \be \vecy(t)=(1-\gamma)\frac{\vecx(t)}{N_x}+\gamma
\frac{B\veceta(t)}{N_\eta} \label{data2} \ee Here, the signal
$\vecx(t)$ contains truly unidirectional information flux and is
generated using AR-models of order $P=5$ for two channels. In general,
an AR-model is defined as \be
\vecz(t)=\sum_{p=1}^PA(p)\vecz(t-p)+\vecxsi(t) \label{AR} \ee where
$A(p)$ are the AR-matrices up to order $P$ and $\xi(t)$ is white
Gaussian noise with covariance matrix $\Sigma$ chosen here to be the
identity matrix.  For computing Granger Causality, the AR model was
fitted with order $P=10$.

All entries of AR-matrices were selected randomly as independent
Gaussian random numbers with $A_{21}(p)=0$ for the signal part
$\vecx(t)$, corresponding to unidirectional flow from the second to
first signal, and $A_{12}(p)=A_{21}(p)=0$ for the noise part
$\veceta(t)$, corresponding to independent sources. Noise was mixed
into channels with a random $2\times 2$ mixing matrix $B$.  Both the
signal part and the mixed noise part ($B\veceta(t)$) are normalized by
the Frobenius norms of the respective data matrices ($N_x$ and
$N_\eta$) and finally added with a parameter $\gamma$ controlling for
the relative strength.  The time constant implicit in the AR-model was
assumed to be 10 ms, and we generated 60000 data points for each
system and channel. This corresponds to a Nyquist frequency of 50 Hz
and to 10 minutes measurement.  We analyzed systems for all $\gamma$
in the range $[0,1]$ with step $0.1$. For each $\gamma$ we analyzed
1000 randomly selected stable systems with both methods and both for
wide band (using all frequencies) and narrow band analysis. For the
narrow band, we used a band of 5 Hz width, centered this band around
the spectral peak of the (known) signal of interest and analyzed only
cases where the band includes at least 60\% of the total power of the
signal of interest.

\begin{figure}[ht]
\centering\resizebox*{8cm}{7cm}{{\includegraphics{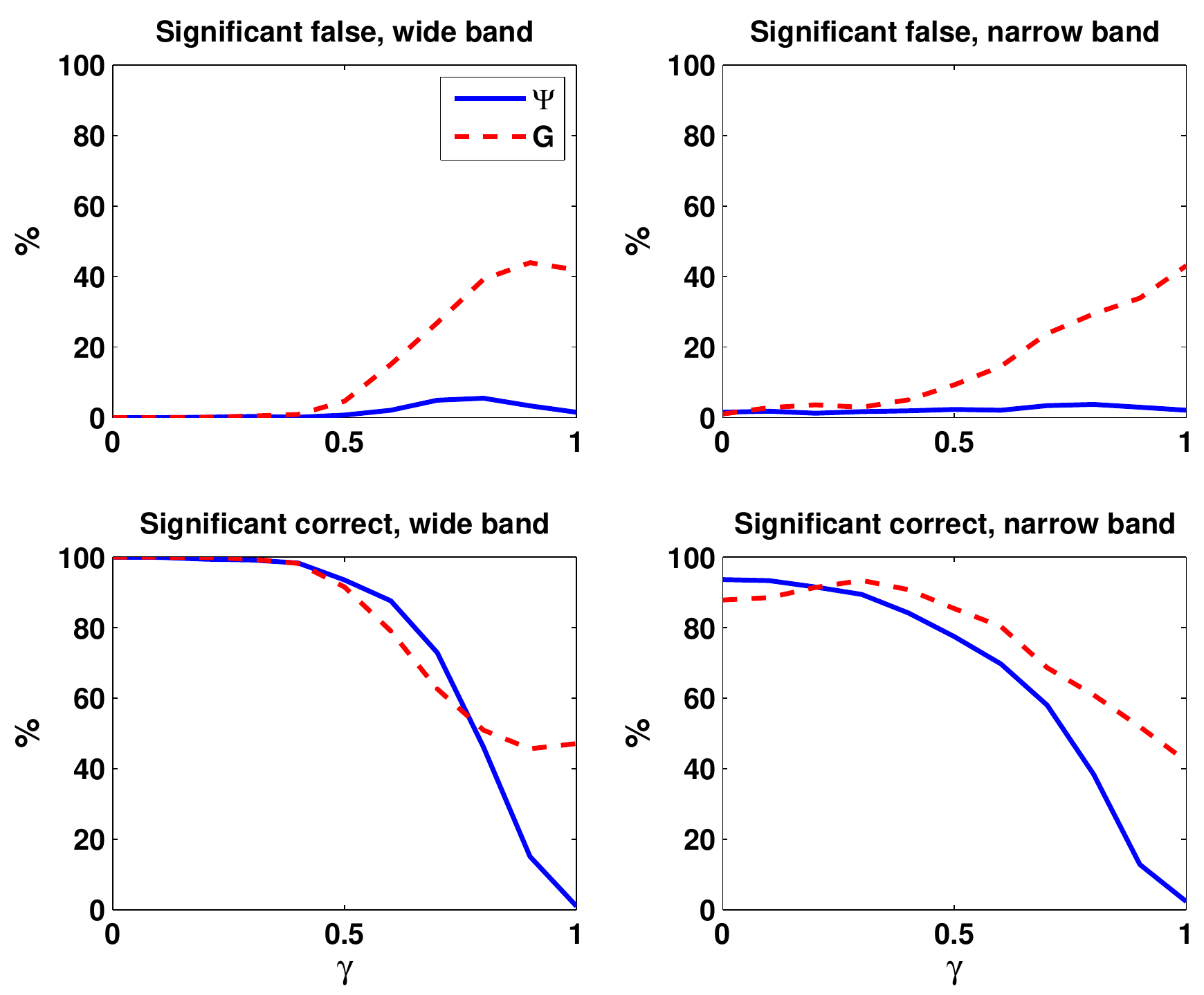}}}
\caption{Fraction of significant detections of Granger Causality and PSI 
as a function of noise level $\gamma$.
\label{fig_simsignificant}}
\end{figure}

The fractions of significant false and significant correct detections
as a function of $\gamma$ are shown in Fig.\ref{fig_simsignificant}.
We observe that for increasing noise level the fraction of significant
false detections for Granger Causality comes close to $50\%$ while PSI
rarely makes significant false detection at all. For PSI, the worst
case observed is at $\gamma=0.8$ for the wide band with $6\%$
significant false detections. This level can be reduced to about
$3.5\%$ if we increase the frequency resolution to $0.25$Hz. However,
the price is some loss in statistical power and it is important to
show that also the proposed method might fail, even if it is unlikely
in the sense of the present simulation.
 
We observe similar significant correct detection rates for both
methods for small and moderate noise levels. For high noise level
Granger Causality shows a much larger fraction of significant correct
detections which, however, is meaningless given the large fraction of
significant false detections.

After having illustrated the robustness of our new method
on simulated data we now apply the PSI to real data, namely EEG.
For this, 88 healthy subjects were recruited randomly by
the aid of the Swedish population register. During the experiment,
which lasted for 15 minutes, the subjects were instructed to relax and
keep their eyes closed. Every minute the subjects were asked to open
their eyes for 5 seconds. EEG was measured with standard 10-20 system
consisting of 19 channels. Data were analysed using linked mastoids reference.  
The protocol was approved by the Hospital Ethics
Committee.

 The most prominent feature of this measurement 
is the alpha peak at around 10 Hz. This rhythm is believed 
to represent a cortico-cortical or thalamo-cortical interaction. 
The direction of this interaction is an open question. While 
it is mostly believed that this rhythm originates in  occipital 
 areas and spreads to other (more frontal) areas \cite{Lop91} this view 
 has also been challenged \cite{Ito05}.

\begin{figure}[htb]
\centering\resizebox*{8cm}{7cm}{{\includegraphics{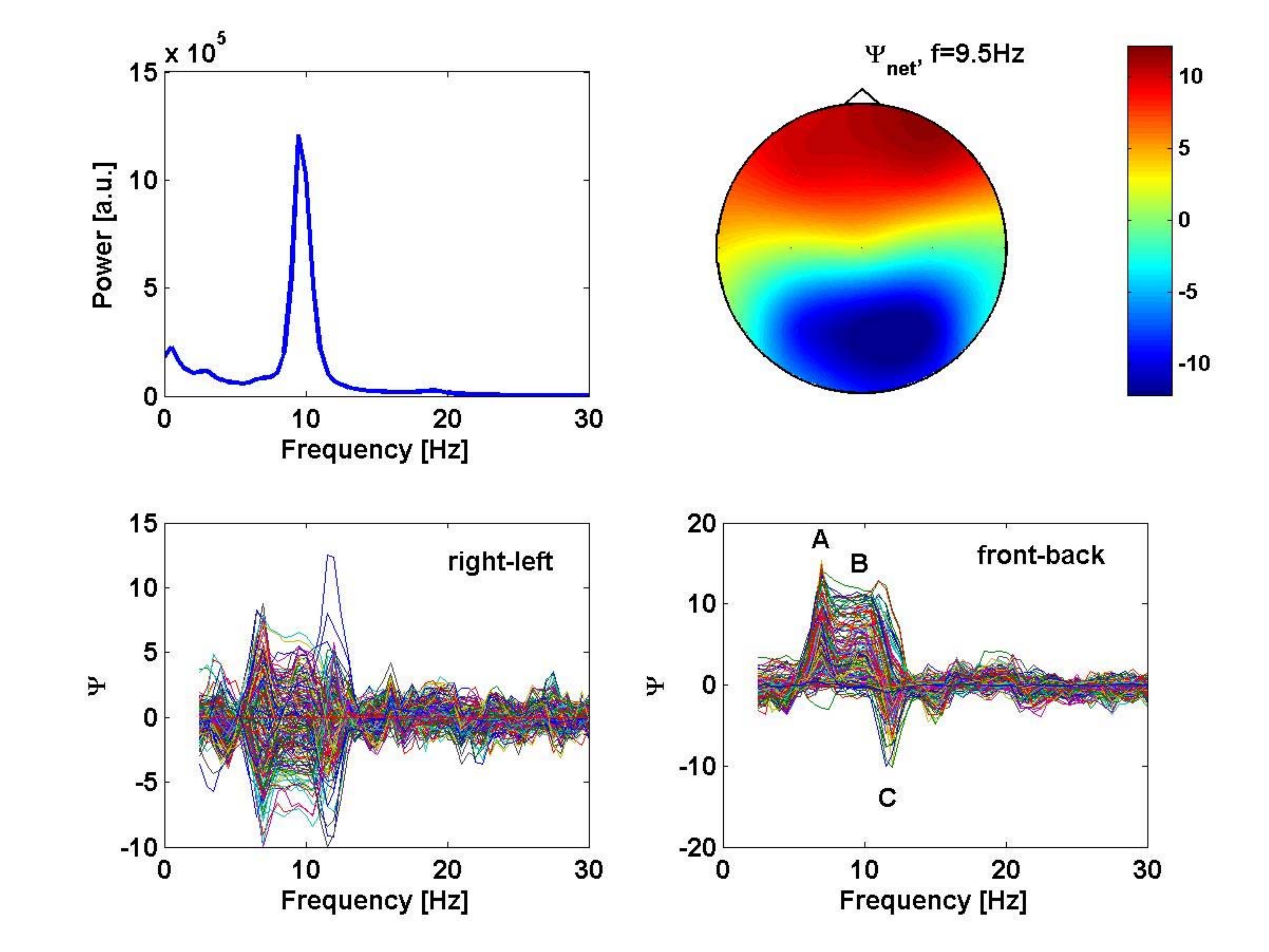}}}
\caption{Upper left: signal power as a function of frequency 
averaged over the two occiptial channels O1 and O2 showing a clear 
alpha peak at $f=9.5$ Hz. Upper right: net information flux at $f=9.5$ Hz.
Lower panels: PSI for all channel pairs and all frequencies projected 
on right-to-left and front-to-back direction, respectively. 
\label{fig_psi_example}
}
\end{figure}

\begin{figure}[htb]
\centering\resizebox*{8.5cm}{7.5cm}{{\includegraphics{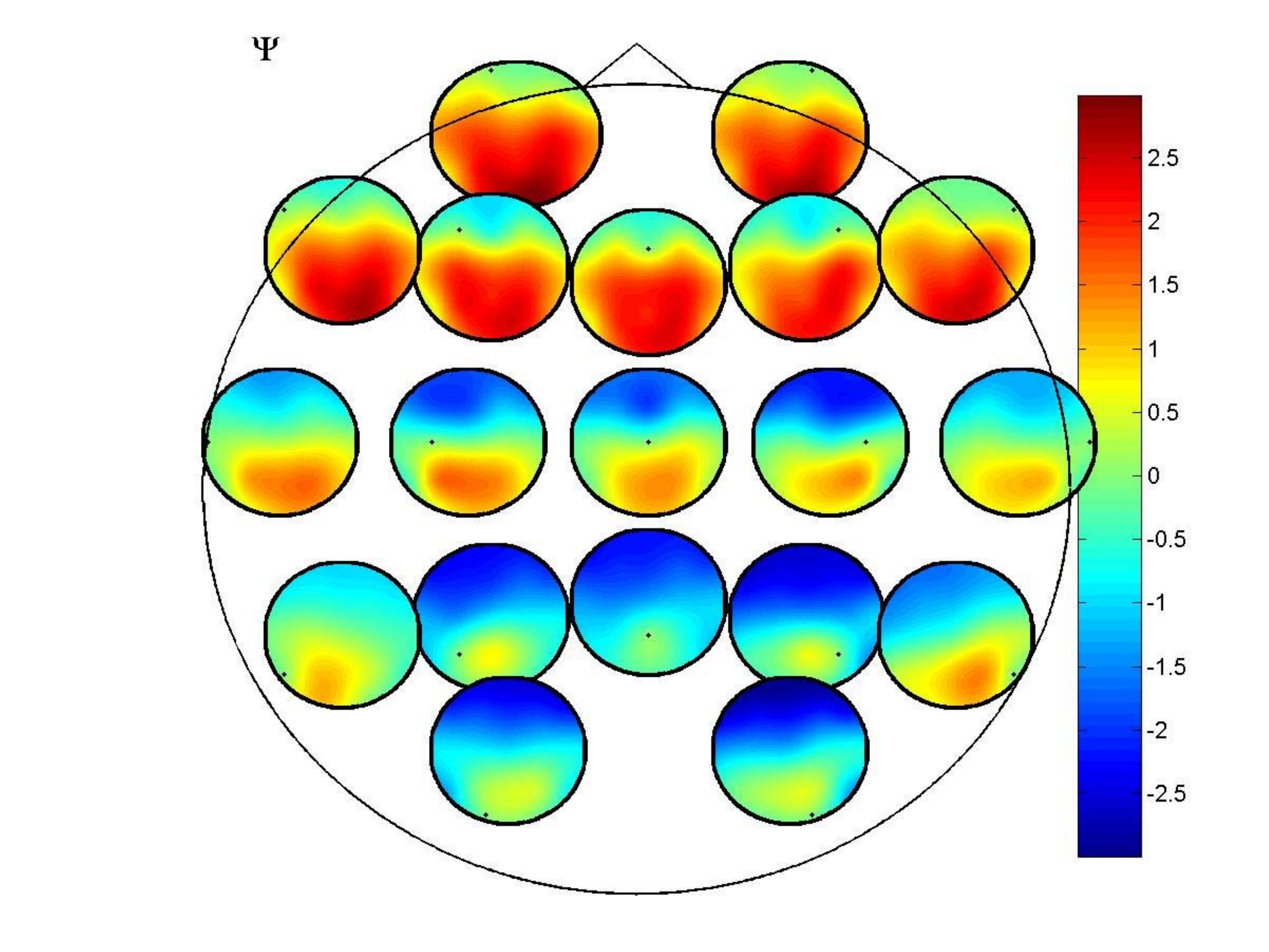}}}
\caption{Phase Slope Index for all pairs of channels averaged over 
all subjects each at the peak of the alpha rhythm. The $i.th$ 
small circle is located at the $i.th$ electrode position 
and is a contour plot of the $i.th$ row of the matrix with elements $\Psi_{ij}$. 
The red color in frontal circles indicates that the frontal electrodes are estimated 
as the drivers.
\label{fig_ps_ga} }
\end{figure}

For illustration we  show results for PSI for one selected subject 
in Fig.\ref{fig_psi_example}. 
The power (upper left panel), averaged over the two occipital channels (O1 and O2), shows a 
very strong peak at $9.5$ Hz. PSI values were calculated for all channel pairs 
with frequency resolution $0.5$ Hz using a frequency band of $5$Hz width centered around 
frequency $f$. In the upper right panel we show  the net information flux at $f=9.5$ Hz   
 defined for the $i.th$ channel by 
\be 
\Psi_{net}(i,f)=\frac{\sum_j \Psi_{ij}(f)}{std(\sum_j \Psi_{ij}(f)}
\label{netflux} 
\ee
We clearly observe that frontal channels are net drivers ( $\Psi_{net} >0$)
and occipital channels net recipients ($\Psi_{net} < 0$). 

To show prefered direction for all 
pairs of channel and for all frequencies we calculate the respective contribution 
to a given direction in the following way: for channels $i$ and $j$ with locations 
$\vecr_i$ and $\vecr_j$ in the two dimensional plane (as shown in the upper right panel)
respectively, we calculate the normalized 
difference vector 
\be
\delta\vecr_{ij}=\frac{\vecr_j-\vecr_i}{|\vecr_i-\vecr_j|}   
\ee
and project it onto the direction of interest, i.e. onto $\vecu=(-1,0)^T$ for right to left 
direction and onto $\vecu=(0,-1)^T$ for front to back direction. We finally calculate the 
contribution of $\Psi_{ij}(f)$ to direction $\vecu$ as 
\be
\Psi_{i,j}(f,\vecu) =\Psi_{ij}(f)\vecu\cdot\delta\vecr_{ij}  
\ee

Results for all channel pairs and for all frequencies are shown for
right-left information flow (lower left panel) and for front-back
information flow (lower right panel).  We do not observe any prefered
direction in the right-left flow. In contrast, the information flow in
front-back direction shows a clear positive plateau at the
alpha-frequency (indicated with letter 'B') meaning that typically the
frontal channels are estimated as the drivers. We also observe a
positive and negative peak (indicated with letters 'A' and 'C') at
frequencies around 7 Hz and 12 Hz, respectively.  Note that these
peaks differ by the width of the frequency band. They are clearly
artefacts due to inadequate settings of the band. Specifically, the
alpha rhythm has a prefered phase (for given channel pair) which is
irrelevant for the estimation of the phase slope unless the alpha-peak
is right at the edge of the frequency band such that the band covers
only half of the system.

We found a similar structure in about 60\% of the subjects.  An
average over all subjects now showing information flux between all
subjects is shown in Fig.\ref{fig_ps_ga}. We also found a substantial
inter-subject variability, both with regard to PSI and actual phase at
the alpha peak. The origin is interesting but so far unclear and goes
beyond the scope of this letter. Note that Granger Causality did not
yield any consistent spatial pattern, presumably for reasons of high false
negative rates similar to the ones observed in Fig.\ref{fig_simsignificant}.

Recent neuroimaging studies have challenged a simple view on a rest
condition by showing a presence of default states in the cortex, which
display complex patterns of neuronal activation
\cite{laufs03,mantini07}. We here show that not only specific areas
are co-activated during rest state, but they also demonstrate at a
gross level a preferential "default" mode of information flow in the
cortex. Importantly, the drivers of such flow are mostly situated in
the frontal areas, from where many top-down attentional influences are
thought to be originated \cite{gilbert07}. In agreement with the above
mentioned imaging studies, our study suggests that the maintenance of
vigilance is a process displaying a coordination of neuronal activity
with well defined drivers and recipients of information flow.

To conclude, we presented a new method to estimate the direction of
causal relations from time series' based on the phase slope of the
cross-spectra. While it is well known that this slope is an indicator
of the direction, the crucial point here is that we defined an average
of the phase slope such that this average is insensitive to arbitrary
mixtures of independent sources with arbitrary spectra.

We verified the claimed properties of the PSI for random linear
systems also showing that the most prominent method to estimate
direction of information flow, Granger Causality, is highly sensitive
to mixtures of independent noise sources.  Additionally, we showed
that in situations with combined unidirectional flow and undirected
noise our method correctly distinguished the two phenomena - in sharp
contrast to Granger Causality. A final application of our method to
real EEG data shows significant and meaningful results from the
neurophysiological point of view and underlines the versatility of our
new method as a universal tool for estimating causal flow in complex
physical systems that consist of mixtures of subcomponents.

{\bf Acknowledgements.} We 
acknowledge partial funding from DFG, BMBF and EU.

\small{\bibliographystyle{unsrt}

\bibliography{causality}}

\begin{thebibliography}{19}
\expandafter\ifx\csname natexlab\endcsname\relax\def\natexlab#1{#1}\fi
\expandafter\ifx\csname bibnamefont\endcsname\relax
  \def\bibnamefont#1{#1}\fi
\expandafter\ifx\csname bibfnamefont\endcsname\relax
  \def\bibfnamefont#1{#1}\fi
\expandafter\ifx\csname citenamefont\endcsname\relax
  \def\citenamefont#1{#1}\fi
\expandafter\ifx\csname url\endcsname\relax
  \def\url#1{\texttt{#1}}\fi
\expandafter\ifx\csname urlprefix\endcsname\relax\def\urlprefix{URL }\fi
\providecommand{\bibinfo}[2]{#2}
\providecommand{\eprint}[2][]{\url{#2}}

\bibitem[{\citenamefont{Granger}(1969)}]{Gra69}
\bibinfo{author}{\bibfnamefont{C.}~\bibnamefont{Granger}},
  \bibinfo{journal}{Econometrica} \textbf{\bibinfo{volume}{37}},
  \bibinfo{pages}{424} (\bibinfo{year}{1969}).

\bibitem[{\citenamefont{Granger}(1980)}]{Gra80}
\bibinfo{author}{\bibfnamefont{C.}~\bibnamefont{Granger}},
  \bibinfo{journal}{Journal of Econ. Dynamics Control}
  \textbf{\bibinfo{volume}{2}}, \bibinfo{pages}{329} (\bibinfo{year}{1980}).

\bibitem[{\citenamefont{Kaufmann and Stern}(1997)}]{Kau97}
\bibinfo{author}{\bibfnamefont{R.}~\bibnamefont{Kaufmann}} \bibnamefont{and}
  \bibinfo{author}{\bibfnamefont{D.}~\bibnamefont{Stern}},
  \bibinfo{journal}{Nature} \textbf{\bibinfo{volume}{388}}, \bibinfo{pages}{39}
  (\bibinfo{year}{1997}).

\bibitem[{\citenamefont{Narayan and Smyth}(2006)}]{Nar06}
\bibinfo{author}{\bibfnamefont{P.}~\bibnamefont{Narayan}} \bibnamefont{and}
  \bibinfo{author}{\bibfnamefont{R.}~\bibnamefont{Smyth}},
  \bibinfo{journal}{Applied Economics} \textbf{\bibinfo{volume}{38}},
  \bibinfo{pages}{563} (\bibinfo{year}{2006}).

\bibitem[{\citenamefont{Marinazzo et~al.}(2006)\citenamefont{Marinazzo,
  Pellicoro, and Stramaglia}}]{Mar06}
\bibinfo{author}{\bibfnamefont{D.}~\bibnamefont{Marinazzo}},
  \bibinfo{author}{\bibfnamefont{M.}~\bibnamefont{Pellicoro}},
  \bibnamefont{and}
  \bibinfo{author}{\bibfnamefont{S.}~\bibnamefont{Stramaglia}},
  \bibinfo{journal}{Physical Review E} \textbf{\bibinfo{volume}{73}},
  \bibinfo{pages}{No.066216} (\bibinfo{year}{2006}).

\bibitem[{\citenamefont{Brovelli et~al.}(2004)\citenamefont{Brovelli, Ding,
  Ledberg, Chen, Nakamura, and Bressler}}]{Bro04}
\bibinfo{author}{\bibfnamefont{A.}~\bibnamefont{Brovelli}},
  \bibinfo{author}{\bibfnamefont{M.}~\bibnamefont{Ding}},
  \bibinfo{author}{\bibfnamefont{A.}~\bibnamefont{Ledberg}},
  \bibinfo{author}{\bibfnamefont{Y.}~\bibnamefont{Chen}},
  \bibinfo{author}{\bibfnamefont{R.}~\bibnamefont{Nakamura}}, \bibnamefont{and}
  \bibinfo{author}{\bibfnamefont{S.}~\bibnamefont{Bressler}},
  \bibinfo{journal}{Proceedings of the National Academy of Sciences of the
  United States of America} \textbf{\bibinfo{volume}{101}},
  \bibinfo{pages}{9849} (\bibinfo{year}{2004}).

\bibitem[{\citenamefont{Sato et~al.}(2006)\citenamefont{Sato, Amaro, Takahashi,
  Felix, Brammer, and Morettin}}]{Sat06}
\bibinfo{author}{\bibfnamefont{J.}~\bibnamefont{Sato}},
  \bibinfo{author}{\bibfnamefont{E.}~\bibnamefont{Amaro}},
  \bibinfo{author}{\bibfnamefont{D.}~\bibnamefont{Takahashi}},
  \bibinfo{author}{\bibfnamefont{M.}~\bibnamefont{Felix}},
  \bibinfo{author}{\bibfnamefont{M.}~\bibnamefont{Brammer}}, \bibnamefont{and}
  \bibinfo{author}{\bibfnamefont{P.}~\bibnamefont{Morettin}},
  \bibinfo{journal}{Neuroimage} \textbf{\bibinfo{volume}{31}},
  \bibinfo{pages}{187} (\bibinfo{year}{2006}).

\bibitem[{\citenamefont{Albo et~al.}(2004)\citenamefont{Albo, Prisco, Chen,
  Rangarajan, Truccolo, Feng, Vertes, and Ding}}]{Alb04}
\bibinfo{author}{\bibfnamefont{Z.}~\bibnamefont{Albo}},
  \bibinfo{author}{\bibfnamefont{G.~D.} \bibnamefont{Prisco}},
  \bibinfo{author}{\bibfnamefont{Y.}~\bibnamefont{Chen}},
  \bibinfo{author}{\bibfnamefont{G.}~\bibnamefont{Rangarajan}},
  \bibinfo{author}{\bibfnamefont{W.}~\bibnamefont{Truccolo}},
  \bibinfo{author}{\bibfnamefont{J.}~\bibnamefont{Feng}},
  \bibinfo{author}{\bibfnamefont{R.}~\bibnamefont{Vertes}}, \bibnamefont{and}
  \bibinfo{author}{\bibfnamefont{M.}~\bibnamefont{Ding}},
  \bibinfo{journal}{Biol. Cybern.} \textbf{\bibinfo{volume}{90}},
  \bibinfo{pages}{318} (\bibinfo{year}{2004}).

\bibitem[{\citenamefont{Nolte et~al.}(2004)\citenamefont{Nolte, Bai, Wheaton,
  Mari, Vorbach, and Hallett}}]{NolBai04}
\bibinfo{author}{\bibfnamefont{G.}~\bibnamefont{Nolte}},
  \bibinfo{author}{\bibfnamefont{O.}~\bibnamefont{Bai}},
  \bibinfo{author}{\bibfnamefont{L.}~\bibnamefont{Wheaton}},
  \bibinfo{author}{\bibfnamefont{Z.}~\bibnamefont{Mari}},
  \bibinfo{author}{\bibfnamefont{S.}~\bibnamefont{Vorbach}}, \bibnamefont{and}
  \bibinfo{author}{\bibfnamefont{M.}~\bibnamefont{Hallett}},
  \bibinfo{journal}{Clin. Neurophysiol.} \textbf{\bibinfo{volume}{115}},
  \bibinfo{pages}{2292} (\bibinfo{year}{2004}).

\bibitem[{\citenamefont{Nolte et~al.}(2006)\citenamefont{Nolte, Meinecke,
  Ziehe, and M\"uller}}]{nolte06}
\bibinfo{author}{\bibfnamefont{G.}~\bibnamefont{Nolte}},
  \bibinfo{author}{\bibfnamefont{F.}~\bibnamefont{Meinecke}},
  \bibinfo{author}{\bibfnamefont{A.}~\bibnamefont{Ziehe}}, \bibnamefont{and}
  \bibinfo{author}{\bibfnamefont{K.}~\bibnamefont{M\"uller}},
  \bibinfo{journal}{Phys Rev E} \textbf{\bibinfo{volume}{73}},
  \bibinfo{pages}{051913} (\bibinfo{year}{2006}).

\bibitem[{\citenamefont{Nunez et~al.}(1997)\citenamefont{Nunez, Srinivasan,
  Westdorf, Wijesinghe, Tucker, Silberstein, and Cadusch}}]{nunez97}
\bibinfo{author}{\bibfnamefont{P.}~\bibnamefont{Nunez}},
  \bibinfo{author}{\bibfnamefont{R.}~\bibnamefont{Srinivasan}},
  \bibinfo{author}{\bibfnamefont{A.}~\bibnamefont{Westdorf}},
  \bibinfo{author}{\bibfnamefont{R.}~\bibnamefont{Wijesinghe}},
  \bibinfo{author}{\bibfnamefont{D.}~\bibnamefont{Tucker}},
  \bibinfo{author}{\bibfnamefont{R.}~\bibnamefont{Silberstein}},
  \bibnamefont{and} \bibinfo{author}{\bibfnamefont{P.}~\bibnamefont{Cadusch}},
  \bibinfo{journal}{Electroencephalogr. Clin. Neurophysiol.}
  \textbf{\bibinfo{volume}{103}}, \bibinfo{pages}{499} (\bibinfo{year}{1997}).

\bibitem[{\citenamefont{Ding et~al.}(2006)\citenamefont{Ding, Chen, and
  Bressler}}]{Din06}
\bibinfo{author}{\bibfnamefont{M.}~\bibnamefont{Ding}},
  \bibinfo{author}{\bibfnamefont{Y.}~\bibnamefont{Chen}}, \bibnamefont{and}
  \bibinfo{author}{\bibfnamefont{S.}~\bibnamefont{Bressler}}, in
  \emph{\bibinfo{booktitle}{Handbook of Time Series Analysis}}
  (\bibinfo{publisher}{Whiley}, \bibinfo{year}{2006}), pp.
  \bibinfo{pages}{437--459}.

\bibitem[{\citenamefont{Marple}(1987)}]{marple87}
\bibinfo{author}{\bibfnamefont{S.}~\bibnamefont{Marple}},
  \emph{\bibinfo{title}{Digital Spectral Analysis with Applications}}
  (\bibinfo{publisher}{Prentice Hall, Englewood Cliffs, NJ},
  \bibinfo{year}{1987}).

\bibitem[{\citenamefont{Schl\"ogl}()}]{biosig}
\bibinfo{author}{\bibfnamefont{A.}~\bibnamefont{Schl\"ogl}},
  \bibinfo{howpublished}{BIOSIG - an open source software library for
  biomedical signal processing, http://BIOSIG.SF.NET}.

\bibitem[{\citenamefont{da~Silva}(1991)}]{Lop91}
\bibinfo{author}{\bibfnamefont{F.~L.} \bibnamefont{da~Silva}},
  \bibinfo{journal}{Electroencephal. and Clin. Neurophys.}
  \textbf{\bibinfo{volume}{79}}, \bibinfo{pages}{81} (\bibinfo{year}{1991}).

\bibitem[{\citenamefont{Ito et~al.}(2005)\citenamefont{Ito, Nikolaev, and van
  Leeuwen}}]{Ito05}
\bibinfo{author}{\bibfnamefont{J.}~\bibnamefont{Ito}},
  \bibinfo{author}{\bibfnamefont{A.}~\bibnamefont{Nikolaev}}, \bibnamefont{and}
  \bibinfo{author}{\bibfnamefont{C.}~\bibnamefont{van Leeuwen}},
  \bibinfo{journal}{Biol. Cybern.} \textbf{\bibinfo{volume}{92}},
  \bibinfo{pages}{54} (\bibinfo{year}{2005}).

\bibitem[{\citenamefont{Laufs et~al.}(2003)\citenamefont{Laufs, Krakow,
  Sterzer, Eger, Beyerle, Salek-Haddadi, and Kleinschmidt}}]{laufs03}
\bibinfo{author}{\bibfnamefont{H.}~\bibnamefont{Laufs}},
  \bibinfo{author}{\bibfnamefont{K.}~\bibnamefont{Krakow}},
  \bibinfo{author}{\bibfnamefont{P.}~\bibnamefont{Sterzer}},
  \bibinfo{author}{\bibfnamefont{E.}~\bibnamefont{Eger}},
  \bibinfo{author}{\bibfnamefont{A.}~\bibnamefont{Beyerle}},
  \bibinfo{author}{\bibfnamefont{A.}~\bibnamefont{Salek-Haddadi}},
  \bibnamefont{and}
  \bibinfo{author}{\bibfnamefont{A.}~\bibnamefont{Kleinschmidt}},
  \bibinfo{journal}{Proc Natl Acad Sci USA} \textbf{\bibinfo{volume}{100}},
  \bibinfo{pages}{11053} (\bibinfo{year}{2003}).

\bibitem[{\citenamefont{Mantini et~al.}(2007)\citenamefont{Mantini, Perucci,
  Gratta, Romani, and Corbetta}}]{mantini07}
\bibinfo{author}{\bibfnamefont{D.}~\bibnamefont{Mantini}},
  \bibinfo{author}{\bibfnamefont{M.}~\bibnamefont{Perucci}},
  \bibinfo{author}{\bibfnamefont{C.~D.} \bibnamefont{Gratta}},
  \bibinfo{author}{\bibfnamefont{G.}~\bibnamefont{Romani}}, \bibnamefont{and}
  \bibinfo{author}{\bibfnamefont{M.}~\bibnamefont{Corbetta}},
  \bibinfo{journal}{Proc Natl Acad Sci USA} \textbf{\bibinfo{volume}{104}},
  \bibinfo{pages}{13170} (\bibinfo{year}{2007}).

\bibitem[{\citenamefont{Gilbert and Sigman}(2007)}]{gilbert07}
\bibinfo{author}{\bibfnamefont{C.}~\bibnamefont{Gilbert}} \bibnamefont{and}
  \bibinfo{author}{\bibfnamefont{M.}~\bibnamefont{Sigman}},
  \bibinfo{journal}{Neuron} \textbf{\bibinfo{volume}{54}}, \bibinfo{pages}{677}
  (\bibinfo{year}{2007}).

\end{thebibliography}

\end{document}